\documentclass[aps,prl,twocolumn,superscriptaddress,showpacs]{revtex4}
\usepackage{amsmath}
\usepackage{graphicx} 

\begin{document}

\title{Quantum fluctuation driven first order phase transition
in weak ferromagnetic metals}

\author{Jason Jackiewicz}
\affiliation{Department of Physics, Boston College, Chestnut Hill, MA, 02467}

\author{Kevin S. Bedell}
\affiliation{Department of Physics, Boston College, Chestnut Hill, MA, 02467}

\date{\today}

\begin{abstract}
In a local Fermi liquid (LFL), we show that there is a line of weak
first order phase transitions between the ferromagnetic and paramagnetic
phases due to purely quantum fluctuations.  We predict that
an instability
towards superconductivity is only possible in the ferromagnetic state.
At $T=0$ we find a point on the phase diagram
where all three phases meet and we call this a quantum triple point (QTP).
A simple application of the Gibbs phase rule shows that only these three
phases can meet at the QTP.  This provides a natural explanation of the
absence of superconductivity at this point coming from the paramagnetic
side of the phase diagram, as observed in the recently discovered ferromagnetic
superconductor, $UGe_{2}$.
\end{abstract}

\pacs{74.70.Tx, 75.30.Kz, 71.10.Ay}

\maketitle

The study of weak itinerant ferromagnetism, both experimentally and theoretically, is an
extremely important topic in understanding strongly correlated electron systems.
It has been shown over the past five years or so
that materials which can be considered weak ferromagnets (small Curie temperature)
display a very wide assortment of complex phenomena and novel physical properties.  For example,
$UGe_2$, $ZrZn_2$, and $URhGe$ have been observed
to be superconducting \textit{and} ferromagnetic, whereas it had previously been expected
(yet not observed) to be
superconducting on only the paramagnetic side of the phase transition \cite{saxena2000,
pfleiderer2001,aoki2001}.
It is even more interesting because this superconductivity might be BCS-like, i.e., singlet
pairing, as opposed to all of the recent models which predict triplet pairing.
The s-wave singlet model has been considered by these authors and others
and there are no definite answers as of yet \cite{karchev2001,blagoev1998,blagoev1999,
abrikosov2001,suhl2001}.
A new result reported recently \cite{pfleiderer2002} shows that the
magnetic transitions in the heavy fermion itinerant ferromagnetic superconductor
$UGe_2$ are of first order, and therefore there does not exist a quantum critical
point as previously thought.

In this Letter we propose an explanation of the observed
first order magnetic transition and superconducting behavior
based on the `Induced Interaction Model' first proposed by Babu
and Brown \cite{babu1973}
\textit{and} the general properties of a Local Fermi Liquid.  As explained
below, this analysis leads to a thermodynamically consistent, first order phase transition
from the ferromagnetic state to the paramagnetic state.
We end by considering some aspects of the superconductivity.

The local Fermi liquid (LFL) was a concept proposed by Engelbrecht and Bedell to look
at normal paramagnetic metals \cite{engelbrecht1995}.
It is a generalization of the LFL proposed by Nozi\`eres while
studying the single impurity
Kondo problem \cite{nozieres1974}.  Blagoev \textit{et al} recently studied
a LFL to explain weak ferromagnetic metals \cite{blagoev1998, blagoev1999}.
This was shown to reproduce
non-trivial results and even the possibility of superconductivity. The
superconductivity was predicted to be s-wave on the ferromagnetic
side, due to the constraint of the LFL, and it had been speculated
to be p-wave on
the paramagnetic side.  While the nature of the superconducting
order parameter is still in question, the real mystery is
why superconductivity is only found in the ferromagnetic
state.

To gain insight into this problem, we start with a simple model
that can describe a strongly correlated Fermi liquid, the LFL of
ref.\cite{engelbrecht1995}.
This theory makes the assumption that the quasiparticle
self-energy $\Sigma(\omega)$ is momentum independent.  This leads
to a further simplification in the theory since only the s-wave
Fermi liquid parameters and scattering amplitudes are nonzero.
In the limit of small magnetic moment, the scattering amplitude
in the LFL can be expressed as
$A^{\sigma\sigma'}_{0}=A_{0}^{s}+A_{0}^{a}\sigma\cdot\sigma'$,
where the $A's$ are the scattering amplitudes, related to the
Landau parameters by
$A_{0}^{s,a}=F_{0}^{s,a}/(1+F_{0}^{s,a})$.
Note that here and elsewhere, the capital
letter quantities
have been made dimensionless by multiplication by the
density of states, $A=N(0)a$ and $F=N(0)f$.  In the LFL
the forward scattering sum rule, which is a consequence
of the Pauli principle, imposes the constraint that
$A^{\uparrow\uparrow}_{0}=A_{0}^{s}+A_{0}^{a}=0$.
It can be shown that when these constraints are applied to
a paramagnetic Fermi liquid, the system is stable against a
transition to a ferromagnetic state and also against phase
separation \cite{engelbrecht1995}, i.e., as $F_{0}{}^{s}$ gets large
then $F_{0}{}^{a}$ saturates to -$\frac{1}{2}$ and superconductivity
in both the s and p-wave channels is suppressed.

These results change dramatically when the LFL is applied
to a weak ferromagnetic system (see refs.\cite{blagoev1998,blagoev1999}
for more details).
In the vicinity
of the phase transition, $F_{0}{}^{a}\rightarrow -1^{-}$ and
$F_{0}{}^{s}\rightarrow -1^{+}$ and both
scattering amplitudes diverge, indicating an instability in the
spin ($A_{0}{}^{a}$) and in the charge ($A_{0}{}^{s}$) sector.
The ferromagnetism remains while
the singlet scattering amplitude,
$A_{0}^{sing}$, is attractive.  This opens up the possibility of the
existence of s-wave superconductivity since the triplet scattering
amplitude is strictly zero in the LFL.

For small magnetization and low temperatures and energies, one can see
that this description remains valid within weak ferromagnetic
Fermi liquid theory.  However, as $m_{0}$ (magnetization) gets even
smaller as one moves toward the critical transition, the situation changes.
Since in the local limit, $m^{*}/m\sim z^{-1}\sim$ log$m_{0}$, around the
critical point the effective mass
diverges, the quasiparticle
residue $z$ goes to zero, and the validity of Fermi liquid theory becomes
questionable.  Hence, the ferromagnetic and paramagnetic LFL states
are not \textit{continuously} connected through the critical point within this
theory.  But, and this is a crucial point,
if the transition is `preempted' by a first-order transition
as to restrict or limit the divergences, then we can recover a consistent theory.
This is indeed what is seen to happen as will be explained
below.

We have shown that the LFL yields unexpected and quite distinct predictions
about the behavior of the paramagnetic and ferromagnetic Fermi liquids.  Moreover,
we argued that they are not connected by a continuous second order phase transition.
What we will show here is that they are in fact connected by a first order transition
using the
`Induced Interaction Model' \cite{babu1973}, which was
further developed by Bedell and collaborators
\cite{ainsworth1987,quader1987}. This a model for self-consistently
calculating the quasiparticle scattering
amplitude (fully reducible interaction) in terms of the
three interaction channels: particle-hole, exchange particle-hole(induced
interaction), and the particle-particle channel.

\begin{figure}
\includegraphics[width=7.5 cm,height=5 cm]{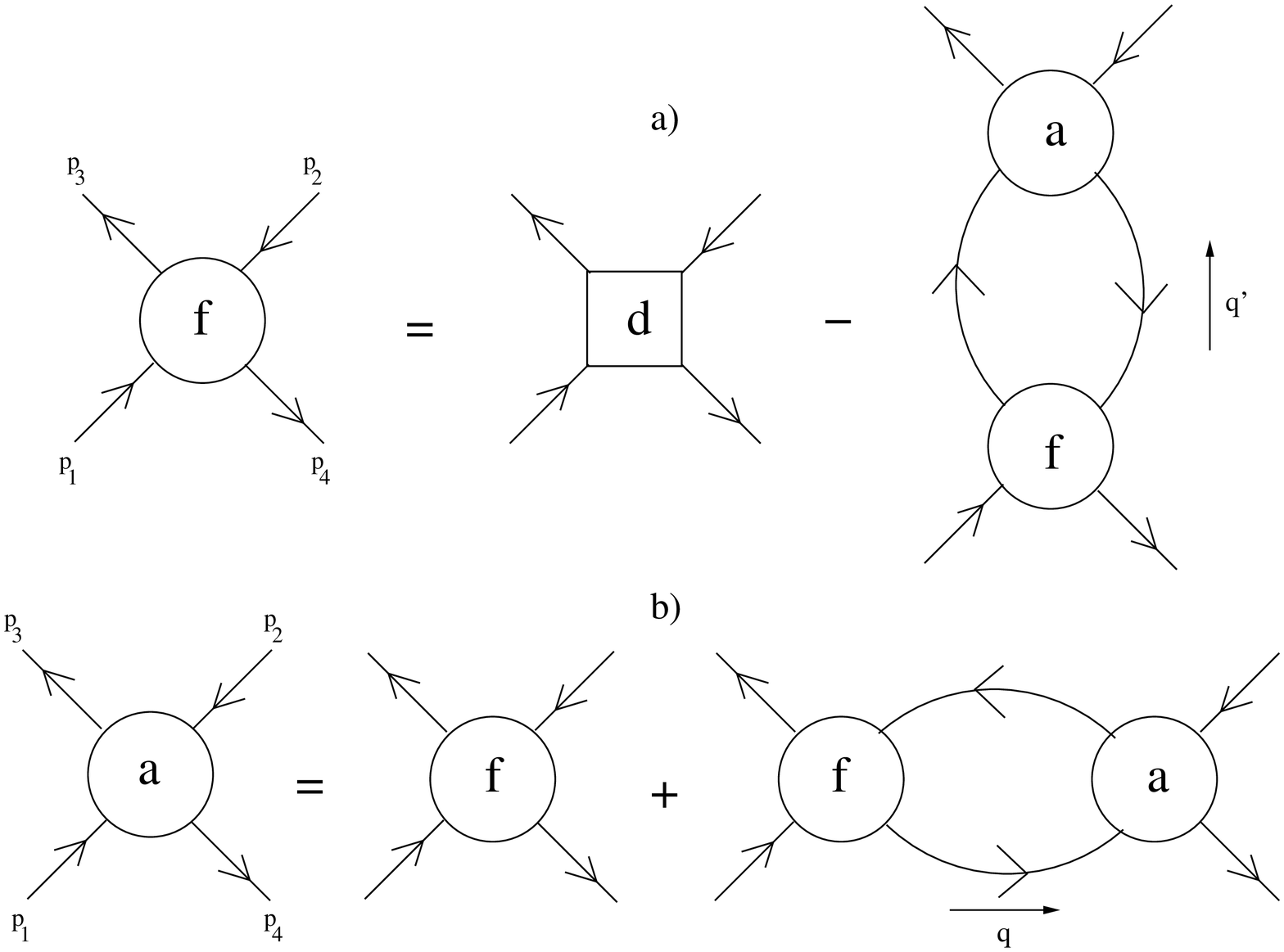}
\caption{The schematic integral equations for the Landau $f$-function and the
scattering amplitudes $a$.  The $\vec{p}$'s are the incoming
and outgoing particle momenta.  Note how the interaction in a) is graphically
shown to be decomposed into a direct term $d$ and the induced term.  Part b)
is the fully reducible set of diagrams.}
\label{induced-fig}
\end{figure}

Including all three channels gives rise to a properly antisymmetrized scattering
amplitude, $a$, where $N(0)a=A$.  The diagrammatic structure of these equations
is shown in Fig.\ref{induced-fig}.  The direct interaction, $d$, is an
antisymmetrized effective two-body potential in the particle-particle
channel.  It is chosen specifically
for a certain physical model, i.e., it contains information about the
underlying Hamiltonian.  The second set of diagrams on the right side
of Fig.\ref{induced-fig}a) is the exchange of topologically equivalent
diagrams in Fig.\ref{induced-fig}b).  Thus, the induced interaction
is a purely quantum effect, arising from the exchange diagrams that are
required to antisymmetrize the effective two-body scattering amplitude.

\begin{figure}
\includegraphics[width=8 cm, height=5 cm]{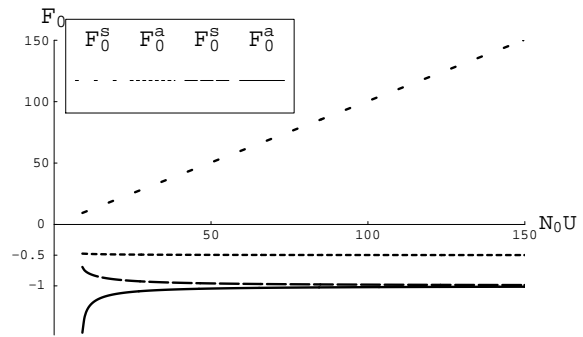}
\caption{The two sets of solutions for the paramagnetic
state (top two lines) and the ferromagnetic state (bottom two lines).
The limiting values for large $N(0)U$ of $F_{0}^{a,s}$ are given in the
text.  Please note the change in scale of the y-axis shown for
clarity.}
\label{fnot-fig}
\end{figure}

To determine the Fermi liquid parameters, we study the equations
of Fig.\ref{induced-fig} in the limit of the momentum transfer,
$\vec{q}=\vec{p_{1}}-\vec{p_{3}}=0$ \cite{ainsworth1987,quader1987}.  In the local
limit of these equations, the induced interactions are equivalent to the
limit of the exchange momentum transfer $\vec{q}'=\vec{p_{1}}-\vec{p_{4}}=0$.  The full
momentum dependence of the interactions on the Fermi surface has
been investigated extensively in the paramagnetic phase \cite{ainsworth1987,
quader1987}, and we are currently extending this approach to the ferromagnetic
phase \cite{jackiewicz2004}.  Including the full momentum dependence will not
change the results we describe here in any qualitative way.  Thus for our
purposes, we will focus here on the local limit of the model to calculate
the quasiparticle interactions for both the ferromagnetic and paramagnetic
LFL.  In the local limit, the coupled \textit{integral}
equations \cite{ainsworth1987,
quader1987} of Fig.\ref{induced-fig}a) and \ref{induced-fig}b) reduce to two coupled
\textit{algebraic} equations,
\begin{eqnarray}
\label{f0s}
F_{0}^{s}=D_{0}^{s}+\frac{1}{2}F_{0}^{s}A_{0}^{s}
+\frac{3}{2}F_{0}^{a}A_{0}^{a} \\
\label{f0a}
F_{0}^{a}=D_{0}^{a}+\frac{1}{2}F_{0}^{s}A_{0}^{s}
-\frac{1}{2}F_{0}^{a}A_{0}^{a}.
\end{eqnarray}

The LFL picture, now coupled with the induced interaction model,
gives a description of a weak ferromagnetic Fermi liquid and
its \textit{first order} transition to the paramagnetic Fermi liquid.
We will take the antisymmetrized direct interaction to be
$D_{0}^{s}=-D_{0}^{a}=N(0)U/2$, where $U$ is the `on-site' contact
interaction, such as in the
Hubbard model.  This is our model-dependent parameter.

We solve eqtn's (\ref{f0s},\ref{f0a}) self-consistently using the
above form of the direct interaction.  We show the results
graphically in Fig.\ref{fnot-fig}.  The different branches for each
solution are described in the caption.  The important consequence is that in the
large $N(0)U$ limit, the solutions remarkably yield exactly the same
results as those of the LFL, namely, $F_{0}^{s}\rightarrow\infty$
and $F_{0}^{a}\rightarrow-1/2$ in one case, and
$F_{0}^{s}\rightarrow-1^{+}$ and $F_{0}^{a}\rightarrow-1^{-}$ in
the other.

Investigating the model more closely we will now employ
certain aspects of a spin-polarized Fermi liquid to the thermodynamics
of the system.  In this theory the Landau parameters
are modified due to a finite magnetization in the ferromagnetic state.
For full details see \cite{bedell1986,castro1989} and references therein.
This has consequences for the effective
masses $m^{*}_{\sigma}$ of each spin species, as well as
for the Fermi momenta $k_{F}^{\sigma}$ of each Fermi
surface, since the magnetization dependence of these quantities
is extremely important.

Since the multiple solutions of this model are found for every
$U$ at large enough $U$, one can imagine the system, for some
specific $U$, jumping from one solution to the other (see Fig.\ref{fnot-fig}).
This can
be shown by examining the chemical potential as a function of
$U$.  We can expand the chemical potential
around a certain $U_{c}$ which determines where one state
gives way to another state of lower energy.  The point
at which the chemical potentials cross is the point of the first order phase
transition. To see this, we calculate the change in the
chemical potential due to the change in the magnetization for
fixed density, $n$, which is given by
$\delta\mu=\frac{1}{4}(C^{\uparrow\uparrow}-C^{\downarrow\downarrow})\delta m$,
where $C^{\sigma\sigma}=1/N^{\sigma}(0)+\tilde{f_{0}}^{\sigma\sigma}$ and
$N^{\sigma}_{0}(0)=k_{F}^{\sigma}m_{\sigma}^{*}/2\pi^{2}$ is the density
of states at the Fermi surface of spin $\sigma$ (the tilde distinguishes the Landau
parameters in the polarized state).
Then the chemical potentials for the two phases near $U_{c}$ are written
as follows:
\begin{eqnarray}
\label{muF}
\mu_{F}(U)\approx\mu_{c}(U_{c})+(U-U_{c})\frac{d\mu_{F}(U=U_{c})}{dU} \\
\label{muP}
\mu_{P}(U)\approx\mu_{c}(U_{c})+(U-U_{c})\frac{d\mu_{P}(U=U_{c})}{dU}.
\end{eqnarray}
\begin{figure}
\includegraphics[width=7 cm, height=4.5cm]{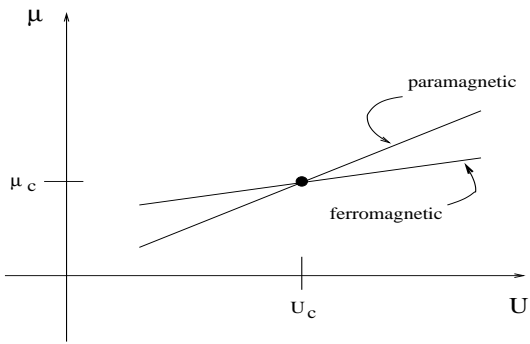}
\caption{The shape of the chemical potential curves expanded
around the critical $U_{c}$.  Note how the higher $U$ regime is the one
in which the ferromagnetic state has the lower energy.}
\label{chempot-fig}
\end{figure}
We differentiate the chemical potential in the ferromagnetic state
implicitly through the magnetization
which is itself a function of $U$. This is seen, for example, through the relation for the
equilibrium magnetization in a weak ferromagnet: $m_{0}\sim\vert1+F_{0}^{a}(U)\vert^{\alpha}$,
where $\alpha$ depends on the order in which the Ginzburg-Landau type expansion
is carried to in the magnetization ($\alpha=1/2$ for maximal $m^{4}$ term, $1/4$ for
$m^{6}$, etc.).  This
holds for small $m_{0}$ \cite{blagoev1998}.  It is seen in Fig.\ref{chempot-fig} that
for smaller $U$ the paramagnetic state is favored but at the critical value $U_{c}$
the chemical potentials cross and the ferromagnetic state is the lower
energy state.

\begin{figure}
\includegraphics[width=7 cm, height=4.8cm]{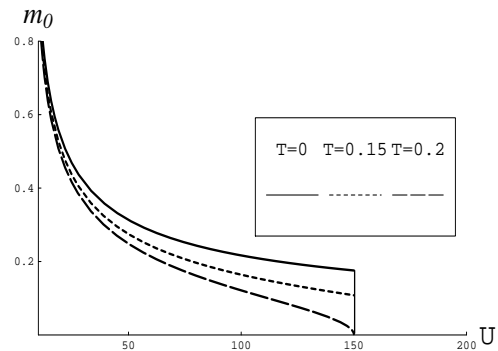}
\caption{The equilibrium magnetization plotted as a function
of the effective interaction $\bar{U}=N(0)U$ for different $T$.  The
temperatures in this model are scaled by the spin fluctuation
temperature, which is about $1/100$ of the Fermi temperature for the
interaction strengths of interest.  The
moment drops discontinuously to zero at the critical $\bar{U}_{c}=150$
at zero temperature and for $T<T_{c}$ (shown by the vertical line),
yet continuously to zero at the critical
temperature $T_{c}=0.2$, indicating a crossover from first order to second
order. See Fig.\ref{pd-fig}.}
\label{mag}
\end{figure}

Extending these calculations to low but finite temperatures
can be done using certain thermodynamic Maxwell relations.  Doing this
we can map out the temperature phase diagram.  The concern will
be with the chemical potential and the pressure at finite
temperatures and small magnetizations where we apply the same
analysis we did at zero temperature.  The details will be
shown elsewhere, but the first step is to integrate the
Maxwell relation,
$-\left (\frac{\partial s}{\partial n}\right )_{T,m}=\left (\frac{\partial\mu}
{\partial T}\right )_{n,m}$,
with respect to $T$, where the entropy density $s$ is given by the
usual low temperature Fermi liquid approximation $s(n,m,T)=\gamma (n,m)T$
\cite{bedell1986}.
This results in a magnetization-dependent chemical
potential expansion in $m$ and up to
second order in $T$.  A similar method is used to develop a
free energy expansion in the magnetization and the temperature.  By
differentiation of the free energy,
an expression for the temperature-dependent magnetization
can be derived.  And finally the pressure, $P=-f+\mu n+Hm$,
can be calculated from the free energy to give an expression in terms
of small magnetization and low temperatures:
\begin{eqnarray}
\label{pressure}
\nonumber
P(m,T)=P(0,0)+N_{0}\frac{\pi^{2}}{6}T^{2}-n\frac{\partial N_{0}}{\partial n}
\frac{\pi^2}{6}T^{2} \\
+G_{1}m^{2}+G_{2}m^4+G_{3}m^{2}\frac{\pi^2}{6}T^{2}
+G_{4}m^{4}\frac{\pi^{2}}{6}T^{2}.
\end{eqnarray}
\noindent The coefficients $G_{i}$ depend on the Landau
interaction functions and the polarization expansion
coefficients of quantities such as the effective mass \cite{jackiewicz2004}.

Now we can determine what happens at temperatures away
from zero.  At zero temperature the magnetization jumps
discontinuously to zero at a certain $U$ indicating a first
order transition.  However, as shown in
Fig.\ref{mag}, when the temperature is turned
on, we see that for a certain critical $\bar{U_{c}}$, and above a certain
temperature $T_{c}\neq 0$, the magnetization goes continuously
to zero, indicating a second order transition.  This implies
that there is a line of first order transitions that ends and
soon becomes a line of second order transitions at finite
temperatures.

\begin{figure}
\includegraphics[width=7 cm, height=4.5cm]{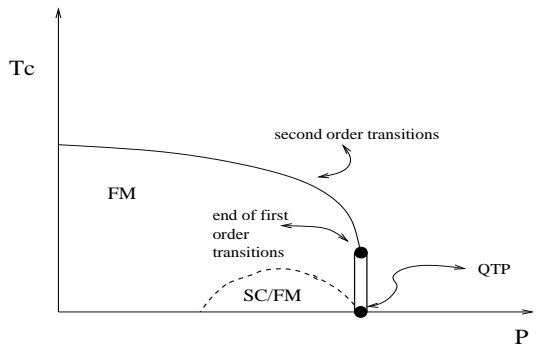}
\caption{The $T_{c}$ versus $P$ schematic phase diagram generated from
this model.  The double line
indicates the line of first order transitions, which ends at a finite $T$,
estimated in the text below.
The SC dome, calculated
from LFL theory only, is taken from ref's \cite{blagoev1998,blagoev1999}.
The superconducting transition
at low temperatures near the QTP is still under
investigation as to its order.  For further discussion on this
point, see \cite{chubukov2003}.}
\label{pd-fig}
\end{figure}

To `translate' from $U$ back to a physical control parameter, such
as pressure, we take advantage of the thermodynamics described above
to develop the phase diagram.  The results are shown in
Fig.\ref{pd-fig}.  The steep slope of the transition near the
critical pressure is due to the fact that the latent heat of the
transition, which is zero at $T=0$, is also zero or nearly zero
at small temperatures.  This is because the entropy difference
between the FM and PM states is proportional to the DOS difference
on each side, which to leading order is zero, and to the next higher order is
$~O(m^{2})$, where $m$ is small.  Thus these transitions are truly
weakly first order.  What we see strikingly resembles
the experimental phase diagram of $UGe_{2}$.  Looking at the
physical values of this system, our effective chosen
$\bar{U_{c}}=150$.  An
approximate calculation of the DOS of $UGe_{2}$ at the Fermi
surface yields a value of $\approx 20/eV$ \cite{bauer2001}.  This
puts $U$ in a range of $10 eV$, a typical value for correlated electron
systems.  Our scaled $T_{c}=0.2$ gives, when a typical Fermi temperature
of $1-10eV$ is used, a critical temperature of around $1meV$, or about
$10K$, the same order of magnitude as seen in the crossover regime
of $UGe_{2}$.

One final thermodynamical observation can be made when looking
at what we call the quantum triple point, QTP.  At this point, as is shown
in Fig.\ref{pd-fig}, three phases end at zero temperature.  According
to the Gibbs phase rule \cite{callen1960}, a single component system, which we
have here, can only accomodate
a maximum of three phases coexisting at a point.  This restrictive
condition explains why superconductivity can only be observed on \textit{one} side
of the QTP, and, as shown in previous studies \cite{blagoev1998,blagoev1999},
this must be the ferromagnetic side, where consequently only pairing in the singlet channel is
attractive.  This is seen experimentally.


In summary, we explicitly have here a microscopic model that
unmistakably yields a first order phase transition from the ferromagnetic to the
paramagnetic state by the inclusion of the \textit{quantum} fluctuations that arise
from the induced interactions. This can be considered the quantum analogue to the case
of first order transitions
driven by \textit{classical} fluctuations in certain liquid crystals studied
by Brazovskii \cite{brazovskii1975}.
Consequently, if we turn off the (quantum) induced terms in our picture,
the model reduces to
a Stoner model which is just a
standard second order transition between the ferromagnetic and paramagnetic phases.
We show that the superconductivity is s-wave and only exists in the ferromagnetic
state, and that, as a consequence of the Gibbs phase rule, there are only three phases that
meet precisely at the quantum triple point.

We gratefully acknowledge discussions with A.V. Balatsky, K.B. Blagoev,
B. Chakraborty, A.V. Chubukov, D. Morr, and especially, Prof. G.E. Brown.
K.S.B. would like to acknowledge the Aspen Center for Physics where some
of the ideas presented here were formulated.
This work was done with the support of DOE Grants No. DEFG0297ER45636 and
No. 60202ER63404.
\bibliographystyle{apsrev}
\bibliography{references}
\end{document}